\documentclass[12pt]{article}
\usepackage{geometry} 
\usepackage{graphicx}
\usepackage{epstopdf}
\usepackage{epsfig}
\usepackage{amsmath}
\usepackage{natbib,hyperref}         
\title{Stock Selection as a Problem in Phylogenetics -- Evidence from the ASX}

\author{Hannah Cheng Juan Zhan$^{1}$,   William Rea$^{1}$,  and Alethea Rea$^{2}$, \\
1. Department of Economics and Finance, University of Canterbury, \\
New Zealand \\
2. Data Analysis Australia, Perth, Australia }

\begin{document}

\maketitle

\begin{abstract}
We report the results of fifteen sets of portfolio selection simulations
using stocks in the ASX200 index for the period May 2000 to December
2013. We investigated five portfolio selection methods, randomly and
from within industrial groups, and three based on 
neighbor-Net phylogenetic networks. We report that using random, industrial
groups, or neighbor-Net phylogenetic networks alone rarely produced
statistically significant reduction in risk, though in four out of the
five cases in which it did so, the portfolios selected using the  phylogenetic 
networks had the lowest risk. However, we report that when using the
neighbor-Net phylogenetic networks in combination with industry group
selection that substantial reductions in portfolio return spread
were achieved.
\end{abstract}
\begin{description}
\item[Keywords: ] Stock selection, ASX200, neighbor-Net networks, portfolio
risk
\item[JEL Codes: ] G11
\end{description}




\section{Introduction}\label{sec:introduction}

Portfolio diversification is critical for risk management because
it aims to reduce
the variance of returns compared with a portfolio of a single stock or similarly
undiversified portfolio. 
The academic literature on diversification is vast,
stretching back at least as far as  \cite{lowenfeld1909}. The modern
science of diversification is usually traced to \cite{markowitz1952}
which was expanded upon in great detail in \cite{markowitz1959}.

In one sense, the approach of \cite{markowitz1952} is optimal and
cannot be improved in the case that, either the correlations and
expected returns of the assets are not time-varying (thus can be accurately
estimated from historical data) or, alternatively, they can
be forecast accurately. Unfortunately, neither of these conditions
hold in real markets leaving the door open to other approaches.

The literature
covers a wide range of approaches to portfolio diversification, such as;
the number of
stocks 
required to form a well diversified portfolio,
which has increased from eight stocks 
in the late 1960's \citep{Evans1968} to over 100 stocks in the late 2000's \citep{Domian2007},
what types of risks should be considered, \citep{Cont2001, 
Goyal2003, Bali2005}, factors intrinsic to each
stock \citep{Fama1992, Fama1993},
the age of the investor, \citep{Benzoni2007}, and
whether international diversification is beneficial, \citep{Jorion1985,Bai2010}, among others.

In recent years a significant number of papers have appeared which
apply graph theoretical methods to the study of a stock or
other financial market,
see, for example, \cite{Mantegna1999}, \cite{Onnela2003a},
\cite{Onnela2003}, \cite{Bonanno2004}, \cite{Micciche2006}, \cite{Naylor2007},
\cite{Kennet2010},
\cite{Djauhari2012}, and 
\cite{Rea2014} among others.

On the pragmatic side,
\cite{DeMiguel2009} list 15 different methods for forming portfolios
and report results from their study which evaluated 13 of these. 
Absent among these 15 methods are any which utilize the above-mentioned
graph theory approaches. This leaves an open question whether 
these graph theory approaches can usefully be applied to the
problem of portfolio selection.

The goal of this paper is to compare three network methods with two
simple portfolio selection methods
for small private-investor sized portfolios.
There are two motivations for looking at very small portfolios sizes.

The first is that,
despite the recommendation of authorites like \cite{Domian2007},
 \cite{Barber2008} reported that in a large sample of 
American private investors
the average portfolio size of individual stocks was only 4.3. 
While comparable data does not appear to be available for
private Australian investors, it seems unlikely that they hold
substantially larger portfolios.  Thus there is a practical
need to find a way of maximising the diversification benefits
for these investors.
The second is that
testing the methods on small 
portfolios gives us a chance to evaluate the potential benefits
of the network methods because the larger the portfolio size, the
more closely the portfolio resembles the whole market and the less
any potential benefit is likely to be discernible.

The mean returns and variances of the individual contributing stocks are insufficient
for making an informed decision on selecting a suite of stocks because 
selecting a portfolio requires an understanding of the correlations between each of the stocks
available for consideration for inclusion
in the portfolio. The number
of correlations between stocks rises in proportion to the square of
the number of stocks, meaning that for all but the smallest of stock
markets 
the very large number of correlations is beyond
the human ability to comprehend them.  \cite{Rea2014} 
presented a method
to visualise the correlation matrix using
nieghbor-Net networks \citep{Bryant2004}, yielding insights into the relationships between 
the stocks.

Another key aspect of stock correlations is the potential change in the correlations with a 
significant change in market conditions (say comparing times of general market increase with 
recession and post-recession periods). 


In this paper we explore investment opportunities on the Australian
Stock Exchange using data from the stocks in the ASX200 index.

Our primary motivation is to investigate five portfolio selection
strategies.  The five strategies are;
\begin{enumerate}
\item picking stocks at random;
\item forming portfolios by picking stocks from different industry groups;
\item forming portfolios by picking stocks from different correlation clusters; 
\item forming portfolios
by picking stocks from the dominant industry group within correlation clusters; 
\item forming portfolios
by picking stocks from non-dominant 
industry groups within correlation clusters.  
\end{enumerate}
Our results show that
knowledge of correlation clusters together with the industry
groups within these clusters can  reduce
the portfolio risk.

The outline of this paper is as follows; Section (\ref{sec:data})
discusses the data and methods
used in this paper, Section (\ref{sec:ASXNNGraphs}) discusses
identifying
the correlation clusters, 
Section (\ref{sec:results}) presents 
the results of the simulations of the portfolio selection methods
and Section (\ref{sec:Discuss}) contains
the discussion and our conclusions.

\section{Data and Methods}\label{sec:data}

We used the  weekly price data for the stocks in the ASX 200 as our dataset. 
Weekly prices along
with the dividend rate and payment date for the period 3 May 2000 to 
4 December 2013
were obtained from DataStream. We appended one or
two letters to each ticker symbol in order to identify the 
industry group for each stock.

Weekly returns were calculated from the price and dividend data
for use in both the portfolio formation simulations and for
estimating the correlations. The correlations were estimated
using the function \verb+cor+ in base R \citep{R} We also calculated
period returns for each stock in each of periods two to six
for use in the simulations.

\begin{figure}[ht]
  \centering
  \includegraphics[width=12cm]{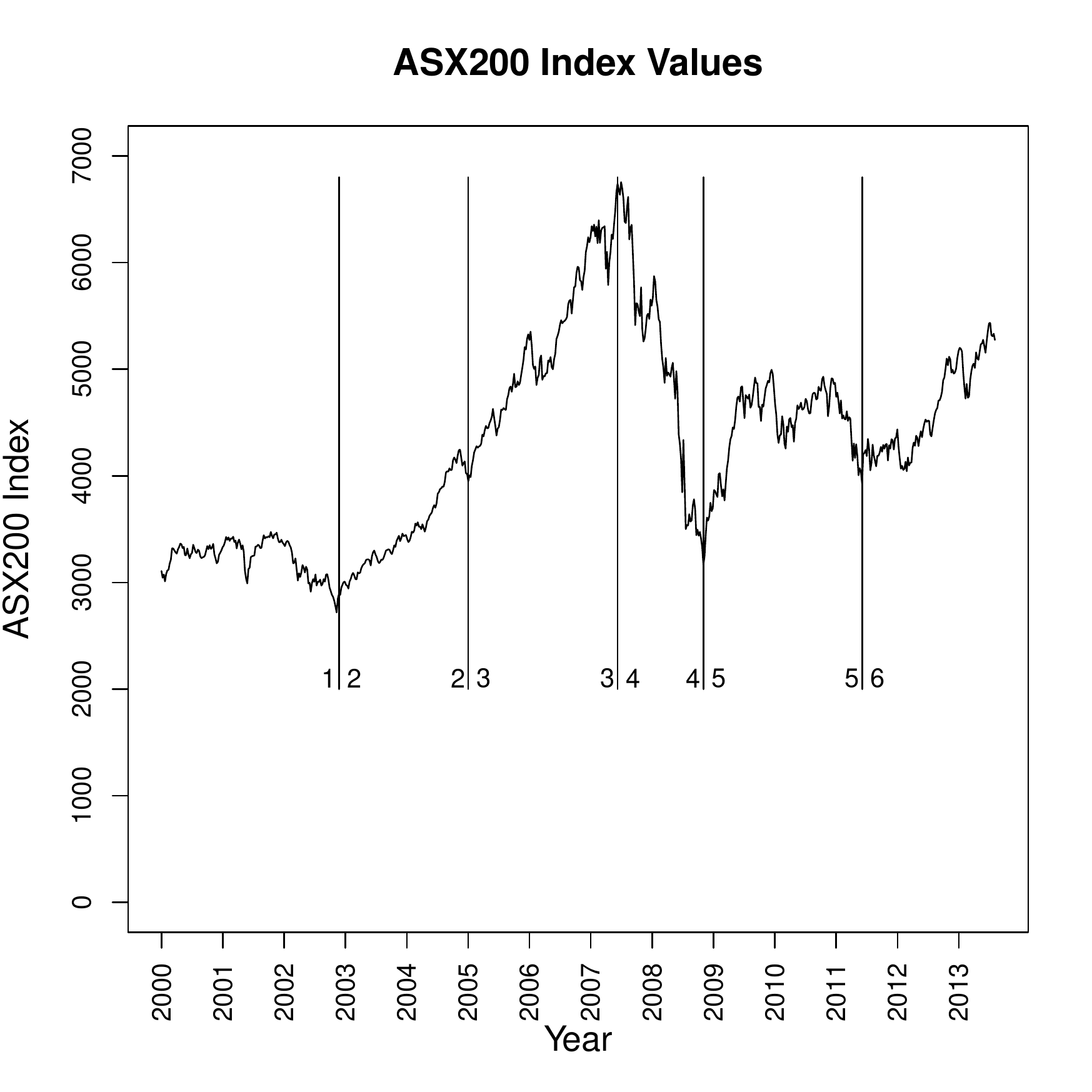}
  \caption{A plot of the ASX200 Index with the
boundaries of the 
study periods marked. 
}
  \label{fig:ASX200Index}
\end{figure}

We divided the whole period into six shorter periods shown in Figure
(\ref{fig:ASX200Index}) and used  out-of-sample testing to test the 
effectiveness of each the five
methods of diversifying portfolios on reducing risk.

\subsection{Neighbor-Net Splits Graphs}

To be able to use clustering algorithms (neighbor-Net is a
clustering algorithm, Appendix (\ref{app:neighbornet}) gives greater 
detail) 
we
need to convert the numerical values in the correlation matrix to
a measure which can be used as a distance. In the
literature the most common way to do the conversion is
by using the so-called ultra-metric given by, 
\begin{equation}
d_{ij}=\sqrt{2(1-\rho_{ij})} \label{eqn:sqrt1mc} 
\end{equation}
where $d_{ij}$ is the distance corresponding to the
the estimated correlation, $\rho_{ij}$, between stocks $i$ and $j$,
see \cite{Mantegna1999} or  \cite{Djauhari2012}
for details.

A typical stock market correlation matrix for $n$ stocks
 is of full rank which means 
that after converting to a distance matrix according to
Equation \eqref{eqn:sqrt1mc}, the location of the points,
here stocks, 
 can only be  fully represented  in  $(n-1)$-dimensional
space. 
In visualization, the high dimensional data space is collapsed
to a sufficiently low dimensional space that the
data can be
represented on 2-dimensional surface such as a page or computer
screen for viewing. Information loss is often unavoidable
in the reduction of the dimension of the data space.  One
of the goals of visualization is to minimize the information
loss while making the structures within the data visible to the
human eye.

Using the conversion in Equation \eqref{eqn:sqrt1mc} we formatted
the converted correlation matrix and 
augmented it with the appropriate stock codes for reading into
the neighbor-Net software, SplitsTree, available from
 \verb+http://www.splitstree.org+ \citep{Huson2006}. 
Using the SplitsTree software
we generated the neighbor-Nets splits graphs. Because the splits
graphs are intended to be used for visualization we defer the
discussion of the identification of correlation clusters and
their uses to Section (\ref{sec:ASXNNGraphs}) 
below.

\subsection{Simulated Portfolios}

Recently \cite{Lee2011} discussed so-called risk-based asset
allocation (sometimes called risk budgeting). 
In contrast to strategies which require both
expected risk and expected returns for each investment opportunity
as inputs to the portfolio selection process, 
risk-based
allocation considers only expected risk. The five methods 
of portfolio selection we
present below can be considered to be risk-based allocation methods.
This probably reflects private investor behaviour in that often they
have nothing more than broker buy, hold, or sell recommendations
to assess likely returns.

The portfolio formation
methods were compared using simulations of 1,000 iterations. There were 
two sets of simulations. 
For first set of simulations a portfolio was sampled based on the rules governing the 
portfolio type using the period return data. 
We recorded the mean and standard deviation of the returns 
for the 1,000 portfolios. The second set of simulations was
carried out in exactly the same manner except weekly return data was
used in order to obtain an estimate of the weekly volatility of the
portfolios. Each set covered  five portfolio formation strategies.

The five portfolio formation strategies are:
\begin{enumerate}
\item Selecting stocks at random;
\item Selecting stocks based on industry groupings;
\item Selecting stocks based on correlation clusters; and
\item Selecting stocks based on the dominant
industry groups within the neighbor-Net  correlation clusters.
\item Selecting stocks based on based on the
non-dominant industry groups within the neighbor-Net correlation clusters.
\end{enumerate}
We describe each of these in turn, combining (4) and (5) into a 
single description.

\begin{description}
\item[Random Selection: ]
The stocks were selected at random using a uniform distribution without 
replacement. In other words each stock was given equal chance of 
being selected according but with no stock being selected twice within a 
single portfolio. 

\item[By Industry Groups: ]
There were 11 industry groups represented among the stocks. 
Some of the groups were very small. For example, the telecommunications 
group only had two representatives in the early 
periods but this increased over time as additional stocks classified as being
in the telecommunications industry were either listed or grew to sufficient size
that they were included in the index. Thus, when the groups were
small, it was necessary to merge
some of them into larger groups for the
purposes of the simulations. This need lessened as the number of stocks
grew. 
We had eight such groups in periods two, three and
four and  nine
groups in periods five and six.

Because the maximum portfolio size was eight stocks 
 the industries were chosen at random using a 
uniform distribution without replacement.
Within each industry group, stocks were selected using
a uniform distribution.

\item[By Correlation Clusters: ]

The correlation clusters were determined by examining the neighbor-Net network 
for the relevant periods (periods one through five). Each stock was assigned 
to exactly one cluster and each cluster can be defined by a single split 
(or bipartition) of the circular ordering of the neighbor-Net for the relevant 
period. The clusters determined in periods one through five  were used to 
generate the stock groups for out-of-sample
testing in periods two through six respectively. Because the
goal of portfolio building is to reduce risk each cluster was paired
with another cluster which was considered most distant from it.
This method is discussed in detail below.

If there were fewer clusters than the desired portfolio size, cluster pairs 
were selected at random and a stock selected from within each correlation 
cluster pair. 
The simulation code was written so that if
 the desired portfolio size was larger than the number of correlation 
clusters 
then each cluster group pair had at
least $s$ stocks selected, where $s$ is the quotient of the portfolio
size divided by the number of clusters. 
Some (the remainder of the portfolio size divided by number of clusters) 
correlation groups will have $s+1$ stocks selected and the cluster
pairs this applied to were chosen using a uniform distribution without
replacement. However, in all cases the number of clusters equalled
or exceeded the number of stocks in the simulated portfolios.


\item[By Dominant or Non-Dominant Industry Group within Clusters: ]

The final two methods relate to selecting stocks from industry groups within 
correlation clusters. Each stock within each  cluster has an associated 
industry group. Therefore each correlation cluster can be subdivided into 
up to eleven sub-clusters based on industry. However, it was clear that
in a number of clusters one industry was dominant, sometimes more
than half of the stocks. This lead us to assign each stock to either the dominant
industry group or the non-dominant industry grouping creating two groups
of stocks within each cluster. This created two disjoint sets of 
clusters with no stock in both groups.

From these two distinct sets of stocks,
simulations were run in the same manner as that
described in ``By Correlation Clusters'' above. These simulations were
not comparable with the three above because the sample sizes were different
and, obviously, each deals with a subset of the data. However, care was
taken to ensure that the sample sizes of both subsets was as close to
equal in size as was practical. 

A problem arose, particularly with the non-dominant industry group
stocks when the number of stocks in the cluster was considered
too small. In these cases we took advantage of the circular ordering
produced by neighbor-Nets and combined the small cluster with a 
neighbouring cluster.

Unchanged from above, each cluster was paired with the one most distant
from it. Once a cluster was selected for inclusion, so was the paired cluster,
and a selection was made using a uniform distribution and, if necessary, without
replacement.

\end{description}
All simulations were coded and run in R \citep{R}.

We used the stocks’ weekly return data in period one to determine the
clusters, then observed period two's return distributions of the simulated
portfolios (1000 replications) which are picked from the different
correlation clusters. Because out-of-sample
testing was used in our analysis, the simulation then was continued for period three, four, five
and six based on the graphs produced from the weekly returns in period two, three, four and
five respectively.




\section{Neighbor-Net Splits Graphs}\label{sec:ASXNNGraphs}

For three of the five types of portfolio simulation methods discussed
above we need to 
identify correlation clusters from the neighbor-Net splits graphs.
In this section we  explain how to identify the clusters, then
proceed to present the neighbor-Net splits graphs and the clusters
identified for each of the first five periods.

At its simplest a neighbor-Net
splits graph is a type of map. The ability to identify correlation
clusters depends on the user's skill in reading it.
As an analogy, all readers of a topographic map 
read the map in the same way. The information the reader extracts
depends on their needs. One person may read a map to extract information
about mountain ranges, another for information on river catchments,
and still another on the distribution of human settlements. But in
all cases  the map readers agree which features are mountains, which
are rivers and which are towns and cities, no confusion arises because
the map is read visually. In the same way all readers of neighbor-Net
splits graphs agree on which feature or features are the origin,
which are splits, which are recombinations and which are the terminal
locations.

Because this is a visual approach, the information extracted
from reading a neighbor-Net splits graph depends on
the researcher or financial analyst
balancing whatever competing requirements they may have.
Here we know that in the simulations to follow
the sizes of the portfolios we
will generate will be two, four, or eight stocks. Consequently,
we do not need large numbers of clusters
and we would like them to have
a sufficiently large number of stocks such that when selecting stocks
at random from within the cluster there are a sufficiently
large number of combinations available to make the simulations
meaningful.
These requirements guide us when identifying clusters in the
neighbor-Net splits graphs. The numbers of clusters and cluster
membership is determined visually and it is important not to confuse
visual with subjective.

\begin{figure}[ht]
  \centering
  \includegraphics[width=14cm]{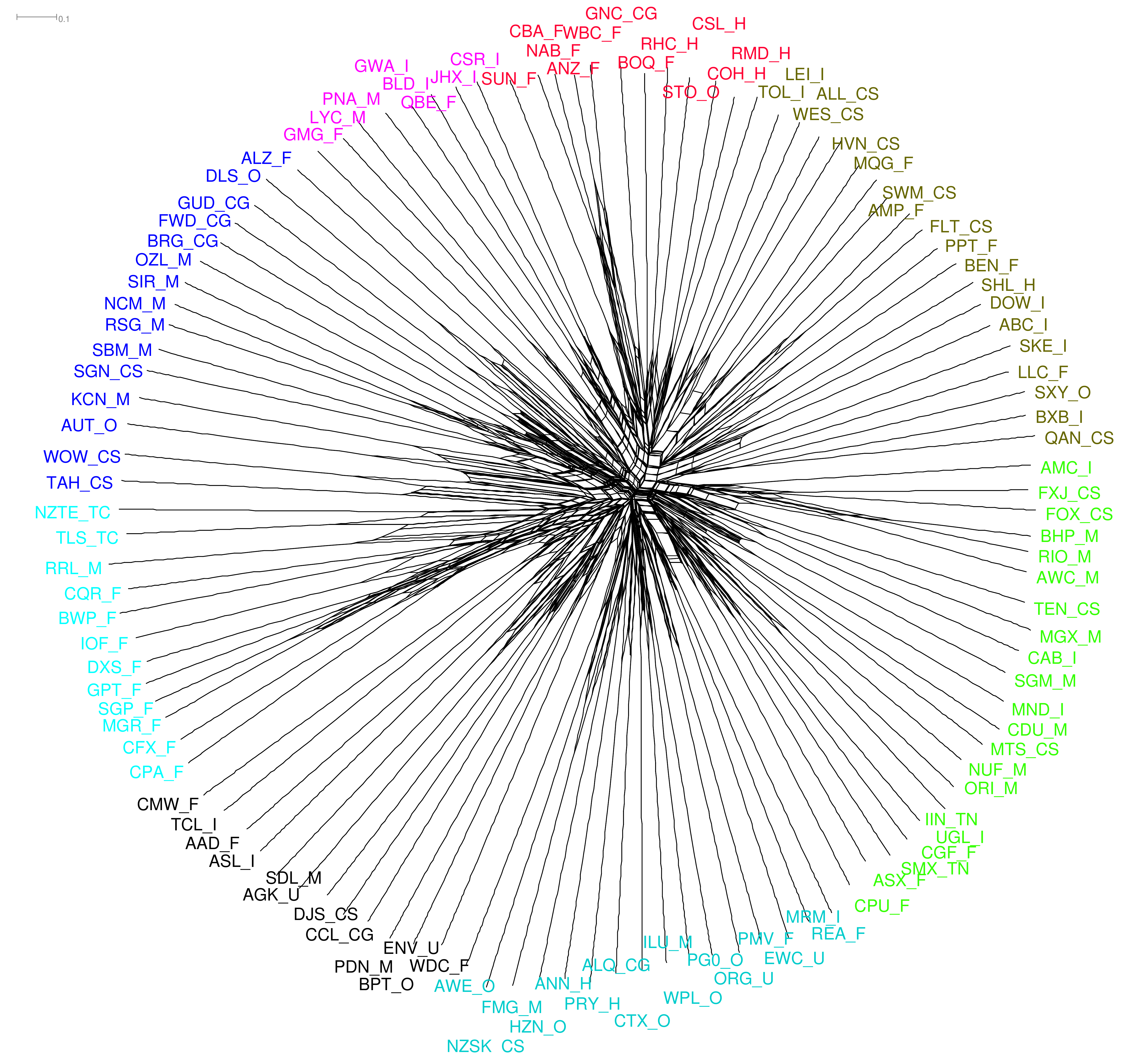}
  \caption{Period 1 neighbor-Nets splits graph with eight clusters identified. }
\label{fig:P1NN}
\end{figure}

\begin{figure}[ht]
  \centering
  \includegraphics[width=14cm]{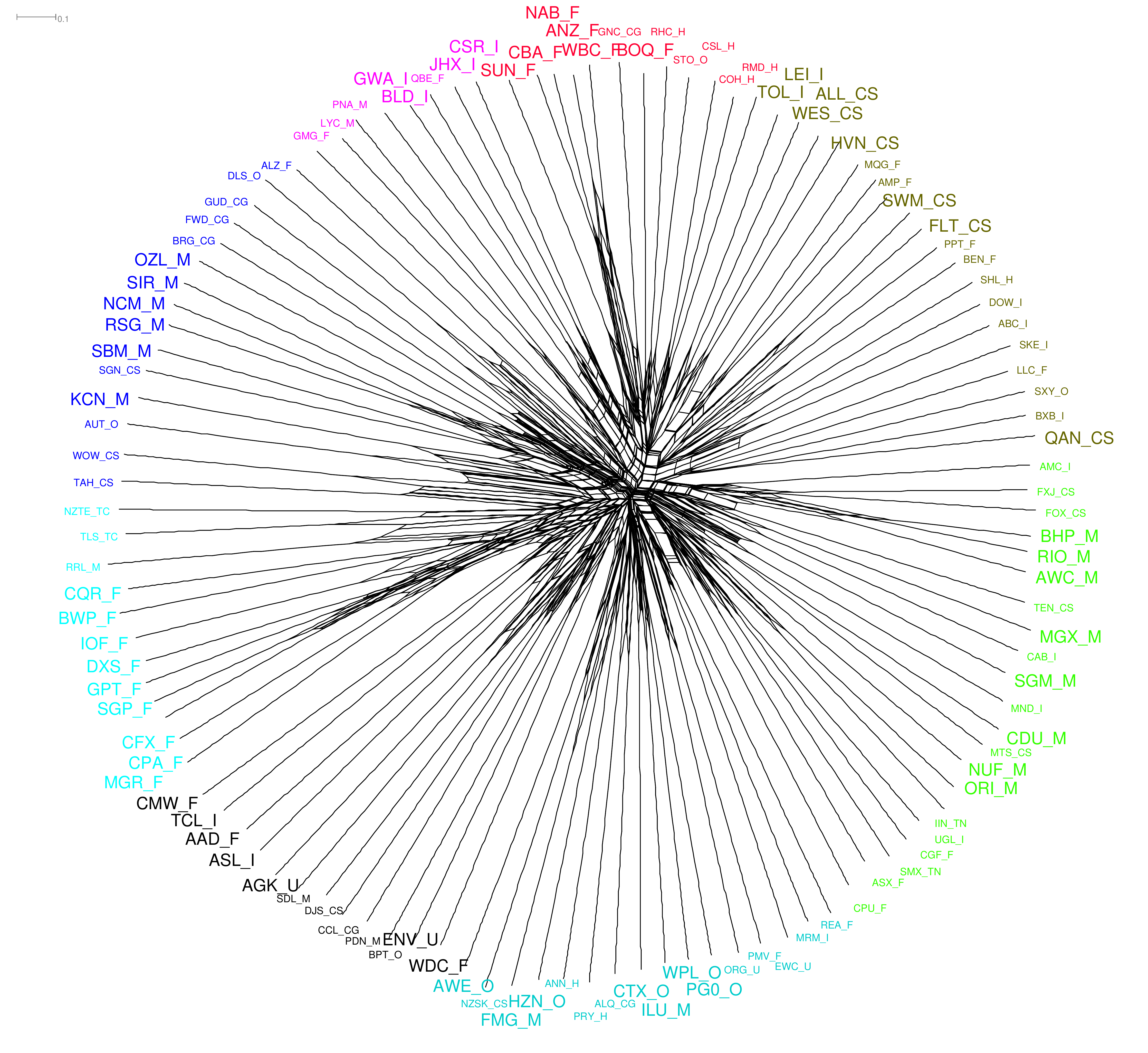}
  \caption{Period 1 neighbor-Nets splits graph with the clusters split
into dominant and non-dominant industries.}
\label{fig:P1NNDomNonDom}
\end{figure}

\begin{figure}[ht]
  \centering
  \includegraphics[width=14cm]{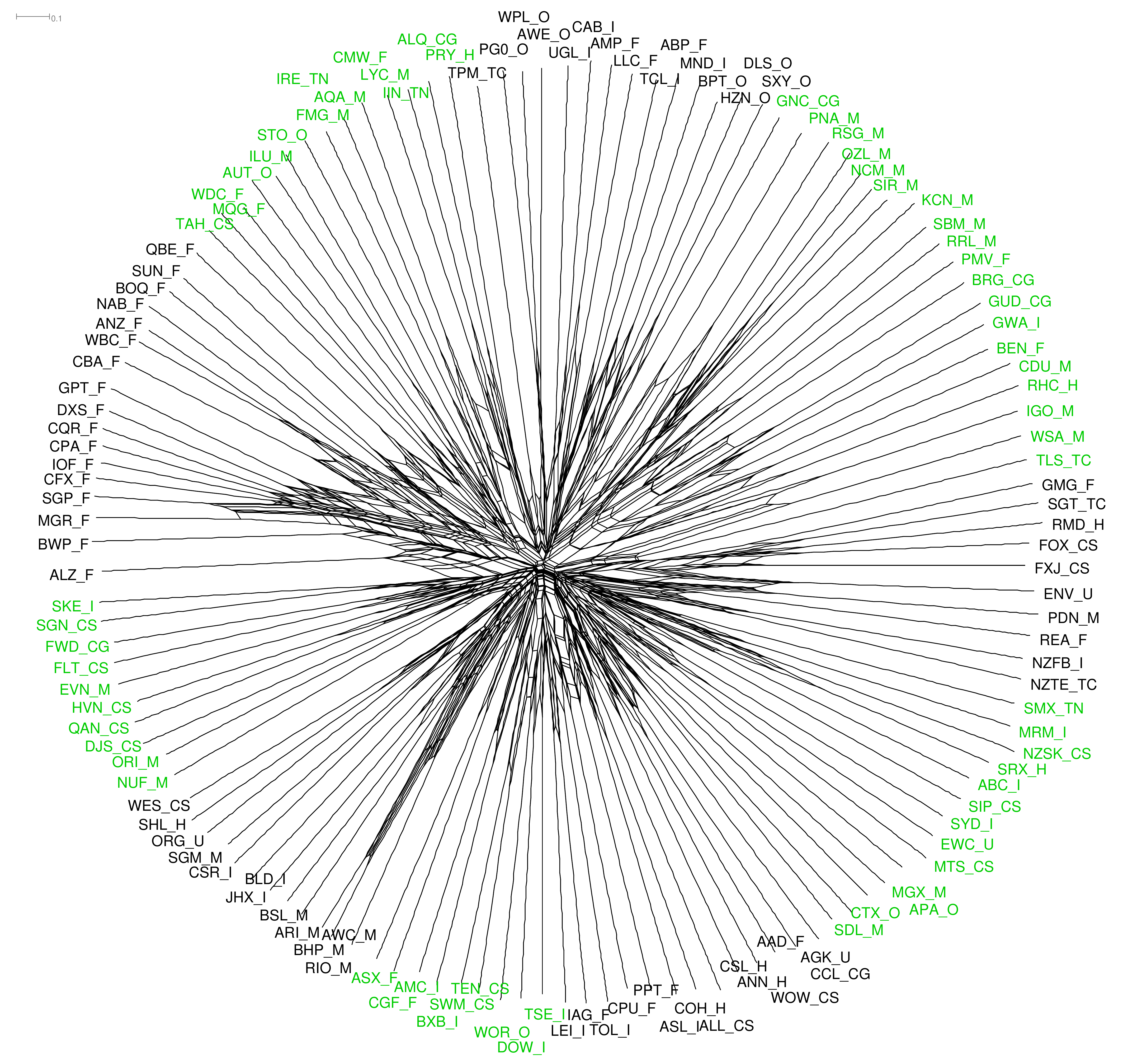}
  \caption{Period 2 neighbor-Nets splits graph with 10 clusters.}
\label{fig:ASXP2Cl10}
\end{figure}

\begin{figure}[ht]
  \centering
  \includegraphics[width=14cm]{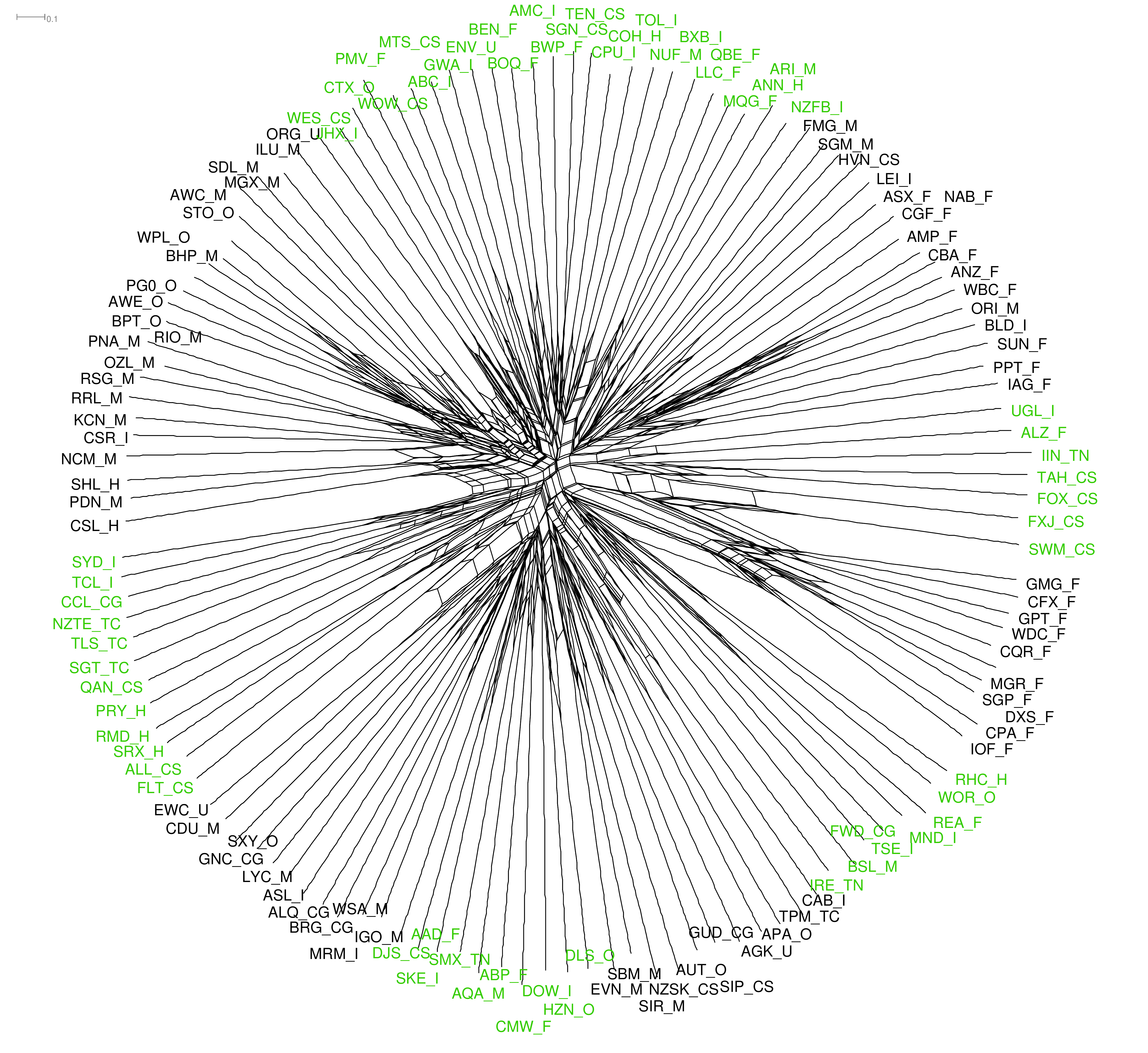}
  \caption{Period 3 neighbor-Nets splits graph with 10 clusters.}
\label{fig:ASXP3Cl10}
\end{figure}

\begin{figure}[ht]
  \centering
  \includegraphics[width=14cm]{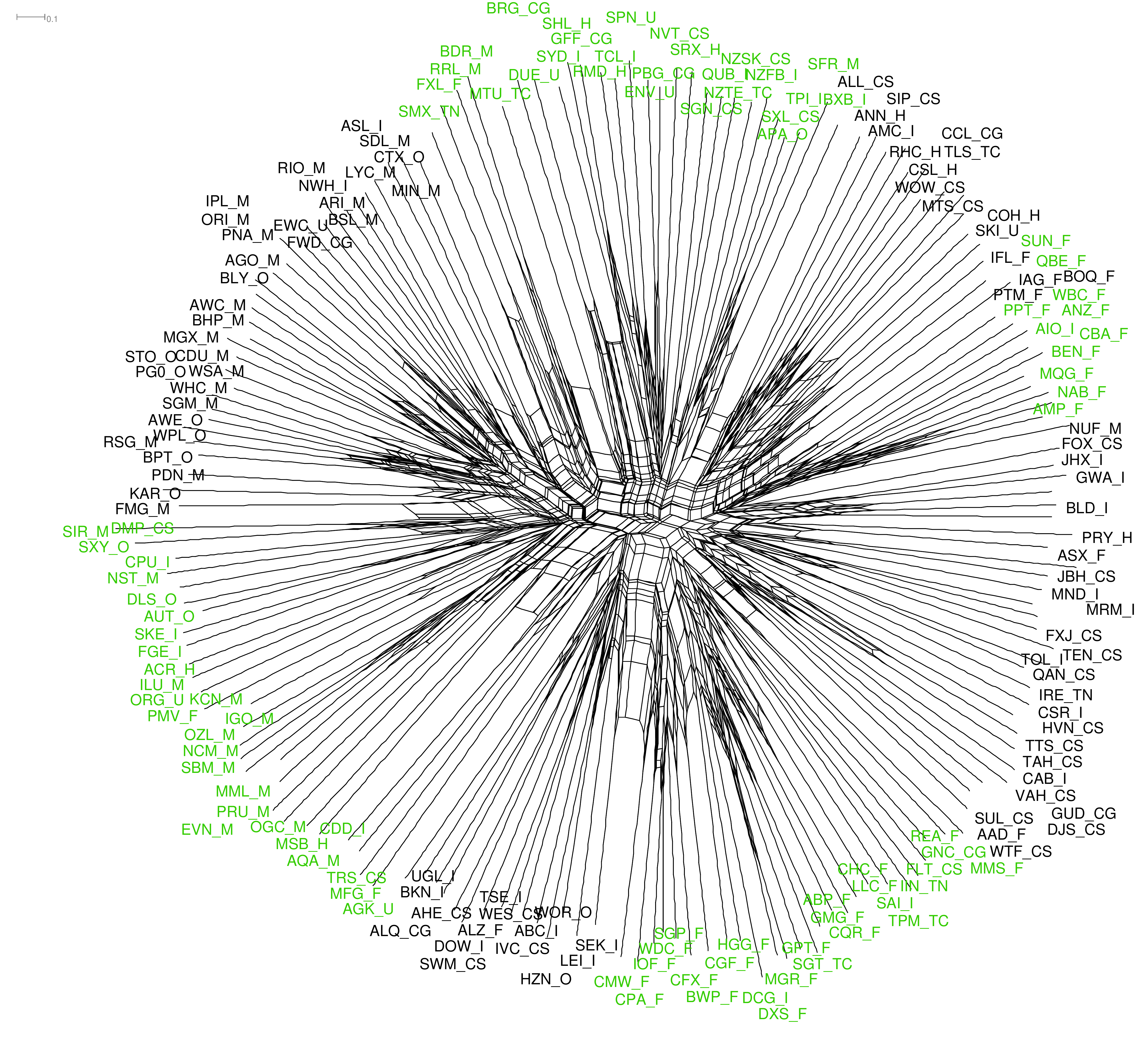}
  \caption{Period 4 neighbor-Nets splits graph with eight clusters.}
\label{fig:ASXP4Cl8}
\end{figure}

\begin{figure}[ht]
  \centering
  \includegraphics[width=14cm]{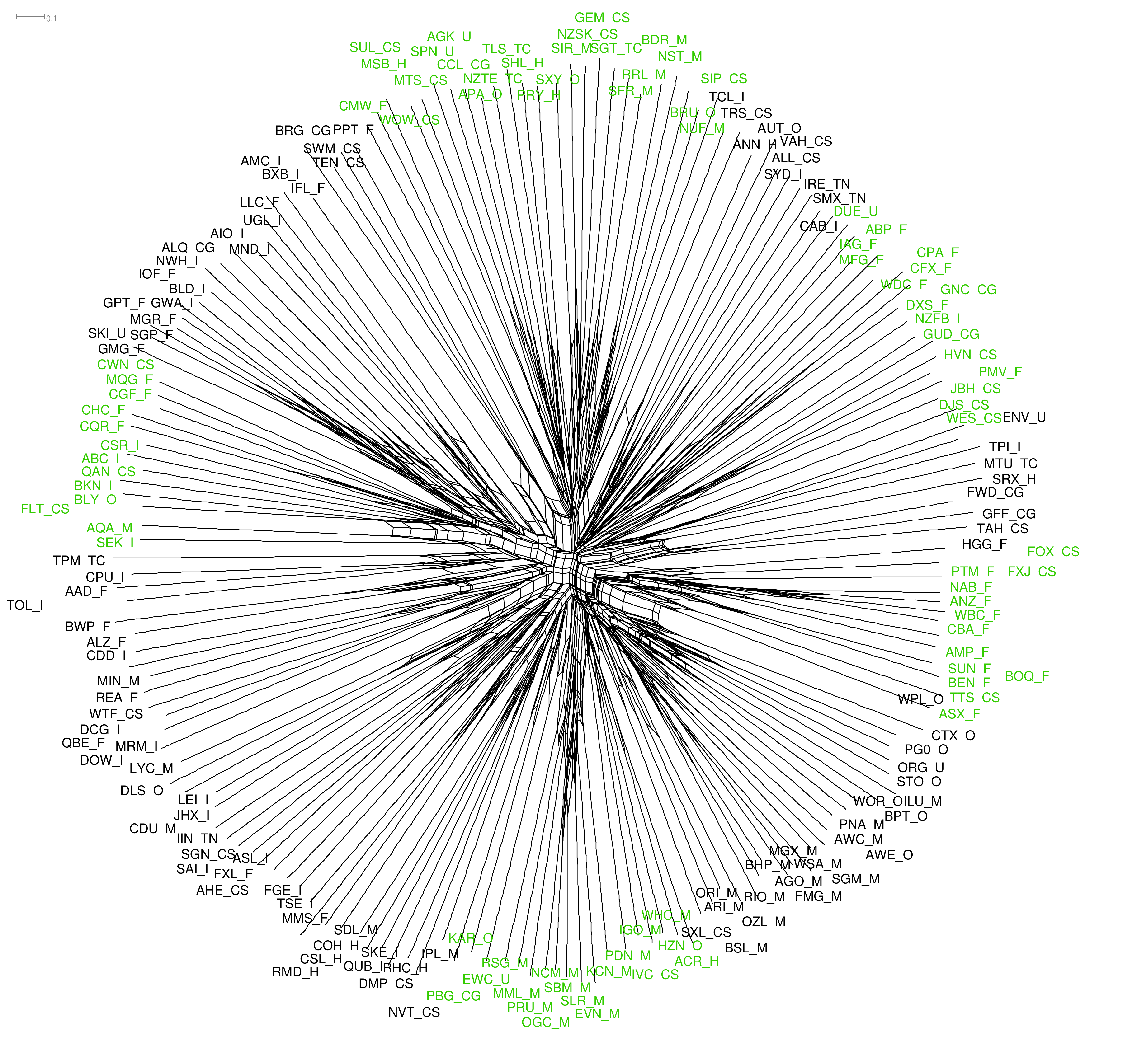}
  \caption{Period 5 neighbor-Nets splits graph with 10 clusters.}
\label{fig:ASXP5Cl10}
\end{figure}

\begin{figure}[ht]
  \centering
  \includegraphics[width=14cm]{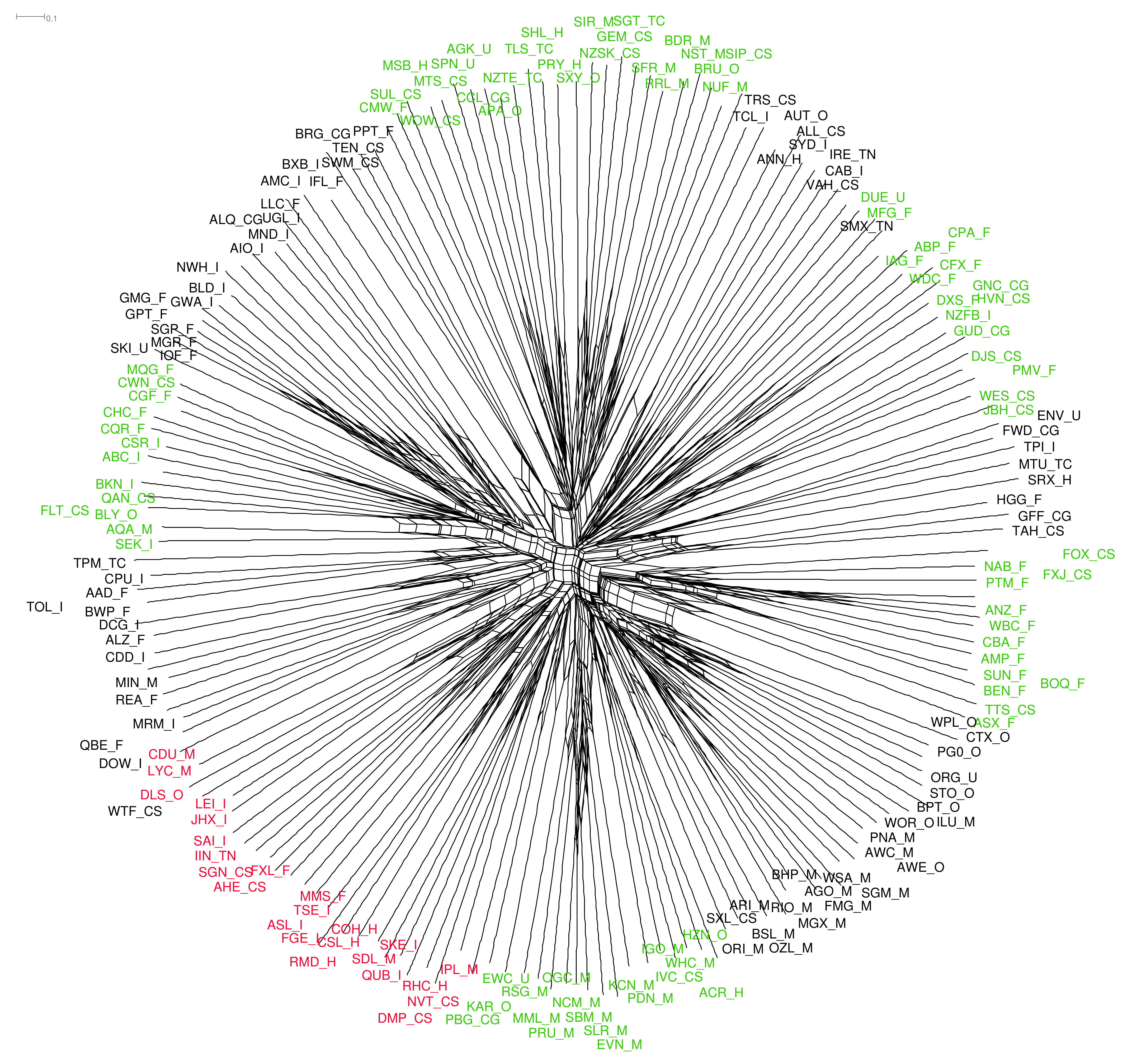}
  \caption{Period 5 neighbor-Nets splits graph with 11 clusters. Cluster
5 in Figure (\ref{fig:ASXP5Cl10}) has been split into two smaller clusters coloured red  and black.}
\label{fig:ASXP5Cl11}
\end{figure}

Figures (\ref{fig:P1NN}) through (\ref{fig:ASXP5Cl11}) present the neighbor-Nets
splits graphs for periods one through five. These were used to determine the
stock groups for out-of-sample testing using data from periods two through
six respectively.

In Figure (\ref{fig:P1NN}) eight clusters were identified and are
colour coded in order to distinguish them. The \verb+SplitsTree+
software generates a large amount of statistical information about
the network. However, we, like other users of neighbor-Nets, look
for breaks in the structure of the network. A good example  can be
found at about eight o'clock in the splits graph between the
stocks labelled CMW\_F and CPA\_F. In its original
context of phylogenetics, if these were species it would tell us that the
last common ancestor was in the distant past. Although CMW\_F and CPA\_F
are placed next to each other in the circular ordering, the two stocks are
not closely related. 

Figure (\ref{fig:P1NNDomNonDom}) presents the same splits graph but within each
cluster the stocks are split into  stocks which belong to the dominant industry
group (larger typeface) and stocks which are in other groups (smaller typeface). 
For example, in the aqua coloured group, the dominant industry is financials
and all but three stocks belong to this sector.

Figure (\ref{fig:ASXP2Cl10}) presents the neighbor-Nets splits
graph for period 2. While the clusters in this figure have not been visually
separated into dominant and non-dominate industry groups,  
the cluster between nine and ten o'clock is dominated by
financial stocks. It is also straight-forward to see
how the clusters were paired. The cluster just mentioned would be
paired with the black coloured cluster between about 2:30 and 3:30 on
the opposite side of the network. These stocks are the most distant
in terms of the circular ordering. If the correlation clusters represent
useful financial groupings of stocks we would expect that choosing a 
pair of stocks from these two clusters would be likely to give a greater
reduction in risk than two stocks selected randomly.

\clearpage 

Figure (\ref{fig:ASXP3Cl10}) gives a good example of where the cluster
pairing may be different. Consider the green coloured cluster at the 
top of the splits graph. It is relatively large and  we shall call it
cluster one. 
At the bottom of
the graph are three  smaller clusters, two black and one
green coloured which we shall call clusters six, seven and eight, reading
the circular ordering in a clockwise direction. The most 
distant cluster from cluster one
(the cluster at 12 o'clock) would be cluster seven (the green coloured cluster 
at six
o'clock). However, we should pair both clusters six and seven
(the black coloured cluster at five o'clock  and
the green coloured cluster at six o'clock respectively) with cluster one.
This illustrates the
fact that while all clusters have a cluster they are paired with, not
all clusters are a reciprocal pair.

Figures (\ref{fig:ASXP5Cl10}) and (\ref{fig:ASXP5Cl11}) are both splits
graphs for period 5 divided into 10 and 11 correlation clusters respectively.
As the number of stocks included in the analysis grows the analyst
gains some flexibility in choosing the number of clusters.

To summarize, a stock market analyst or portfolio manager looks for 
breaks in the structure of the neighbor-Net network when dividing the
stocks into correlation clusters. The \verb+SplitsTree+ software has
considerable flexibility to magnify sections of the network to aid
in decision making which
cannot be easily captured in the static pdf file outputs included 
in this paper. The circular ordering can be very useful when splitting
a correlation cluster into its component industry groups because if 
one or more of the resulting groups are too small to be useful they
can be joined with groups next to them in the circular order. 

\clearpage


\section{Results}\label{sec:results}

%
%
%

\begin{figure}[ht]
  \centering
  \includegraphics[width=14cm]{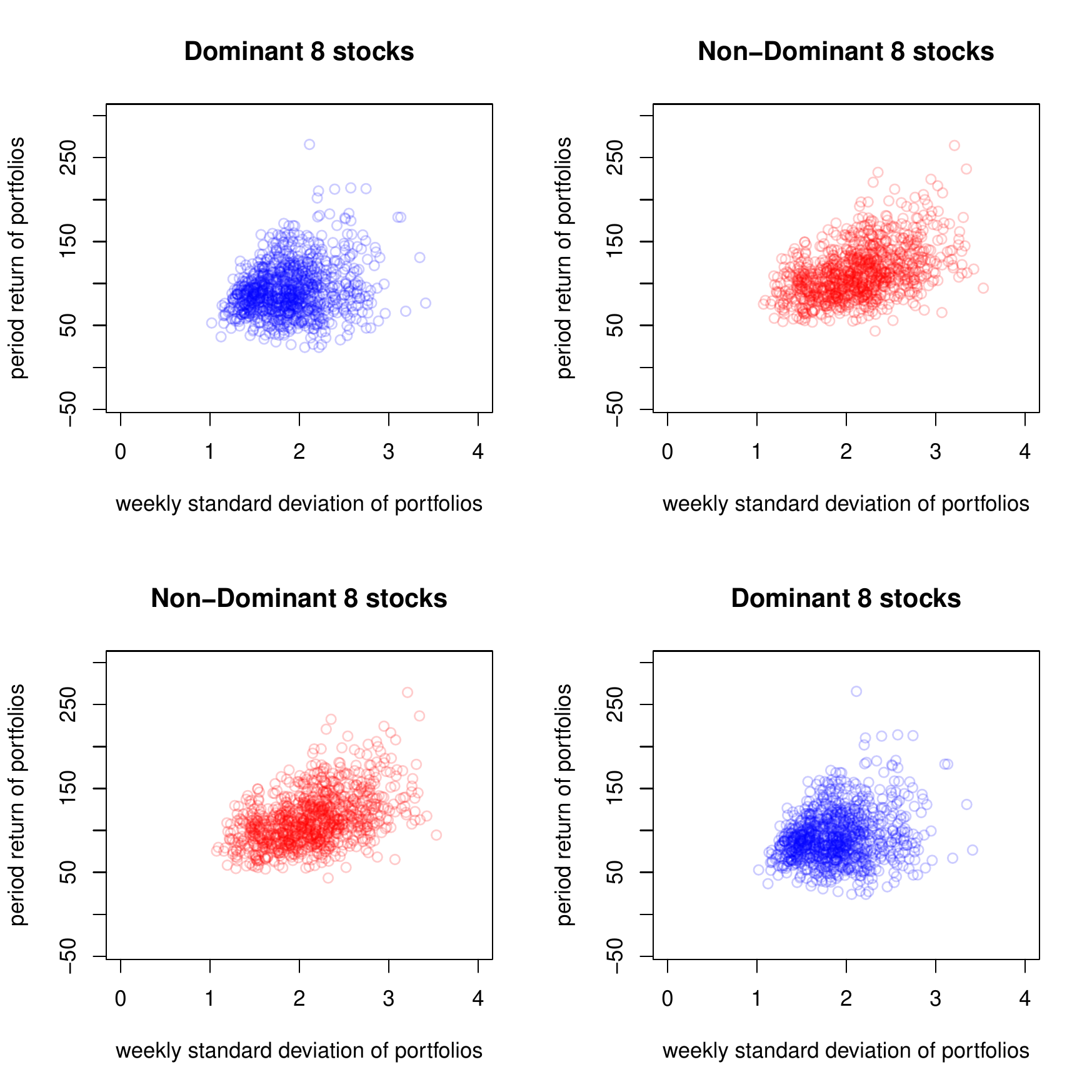}
  \caption{Scatter plots of Period 2 returns against the weekly
volatility for the eight-stock portfolios selected on the basis of
the clusters identified in  neighbor-Nets splits graph from
Period 1 with the clusters split into dominant and non-dominant
industry groups. }
\label{fig:P2Scatter8Stock}
\end{figure}

\begin{figure}[ht]
  \centering
  \includegraphics[width=14cm]{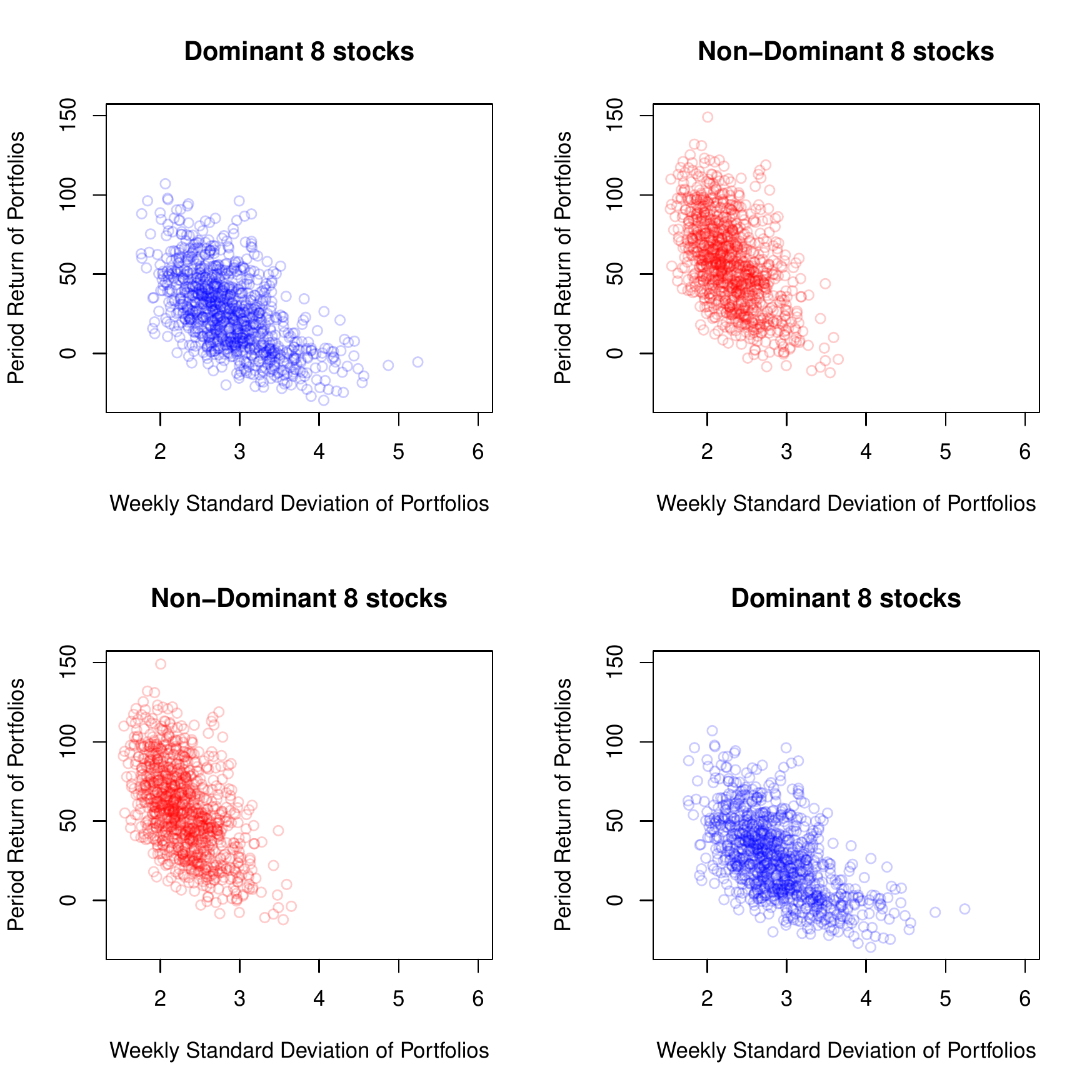}
  \caption{Scatter plots of Period 6 returns against the weekly
volatility for the eight-stock portfolios selected on the basis of
the clusters identified in  neighbor-Nets splits graph from
Period 5 with the clusters split into dominant and non-dominant
industry groups. }
\label{fig:P6Scatter8Stock}
\end{figure}

Tables (\ref{tab:period2}) through (\ref{tab:period6}) together with
Figures (\ref{fig:P2Scatter8Stock}) though (\ref{fig:P6Scatter8Stock})
present the results of the portfolio selection simulations.

In the third part of each table we have labelled the results presented
there the ``Sharpe Ratio'' though this is clearly not the ratio
of \cite{Sharpe1964}. The Sharpe ratio is properly applied to single
assets or single portfolios and estimates the reward to risk ratio.
Here were are dividing the mean portfolio return by the standard deviation
of the returns estimated from the replications. A higher ratio indicates
either a higher mean return or a lower spread of returns generated by
that selection method or some combination of both. We believe the
results are worth reporting but do not discuss them further in this
paper.

\begin{table}
\begin{tabular}{lrrr|rr}
Period 2 &         & Neighbor-Net’s & Industry & Correlation & Correlation  \\
Simulation & Random  & correlation 
 &  Group & cluster with   & cluster without  \\
results  &        & cluster &   &  industry group   &  industry group \\
\hline
Mean return  & &     &   &                   & \\
(2-stock) & 106.49 & 94.93  &  98.52 &  101.80 & 97.24  \\
(4-stock) & 101.51 & 98.45  &  96.93 &   97.55 & 97.12 \\
(8-stock) & 104.66 & 96.90  & 100.82 &  98.892 & 96.14 \\
\hline
Std. Dev.  & &     &   &                   & \\
(2-stock) & 79.34 & *70.51 & 73.87 & 81.90 & *62.47 \\
(4-stock) & *49.39 & 51.85 & 48.96 & 53.14 & *43.62 \\
(8-stock) & *33.61 & 35.64 & 36.56 & 42.33 & *29.81 \\
\hline
Sharpe Ratio  & &     &   &                   & \\
(2-stock) & 1.34        &       1.35  & 1.33 & 1.24 & 1.56  \\
(4-stock) &        2.06 & 1.90        & 1.98 & 1.84 & 2.23  \\
(8-stock) &        3.11 & 2.72        & 2.69 & 2.70 & 3.23  \\
\hline
Levene Tests  & &     &   &                   & \\
(2-stock) & & & 0.001 & &$1.0\times 10^{-7}$ \\
(4-stock) & & & 0.30 & & $1.7\times 10^{-6}$ \\
(8-stock) & & & 0.13 && $1.3\times 10^{-12}$ \\
\hline
\end{tabular}
\caption{Period 2 portfolio performances with standard deviation for
three different portfolio sizes. The selection method with the lowest
standard deviation is marked with an asterisk (*).
The neighbor-Nets method is out-of-sample
testing using the correlation clusters determined from Period 1 data.
The first Levene test is the p-value whether the standard deviation of the
three different portfolio selection methods are equal. The second
result is for the neightbor-Net clusters split into dominant and
non-dominant industry groupings.} \label{tab:period2}
\end{table}

Table (\ref{tab:period2}) presents the results for the Period 2 simulations
which was a period of strongly rising equity prices. The model building
period (Period 1) was a period when the market largely tracked 
sideways with a small decline over the period. Thus the out-of-sample
test represents a strong test of stock selection methods between
the two periods which do not resemble each other.

We are primarily concerned with reducing the risk of the
portfolios. The results in the table give the standard deviation
of the returns of the 1000 replications of the portfolio selection
method. A lower standard deviation indicates that the returns of
the portfolios were more concentrated about the mean portfolio
return. The Levene tests show that only the two stock portfolios
had a significantly different spread of returns. The difference
between the random selection method and neighbor-Nets selection
was almost nine percentage points different.

The final two columns of Table (\ref{tab:period2}) present
the results for the simulations in which the correlation
clusters were divided into dominant and non-dominant 
industry groups. In this case the correlation clusters of
non-dominant industries showed statistically significantly
lower levels of spread of the portfolio returns.

Figure (\ref{fig:P2Scatter8Stock}) plots the returns of
the weekly standard deviation of the portfolio returns,
a measure of volatility, against the period portfolio return for
the eight-stock portfolios dividing the stocks in to dominant
and non-dominant industry groups.
The differences are not pronounced but the spread of returns
is smaller for the correlation clusters with non-dominant 
industry groups though the weekly volatility appears comparable.

\begin{table}
\begin{tabular}{lrrr|rr}
Period 3 &         & Neighbor-Net’s & Industry & Correlation & Correlation  \\
Simulation & Random  & correlation 
 &  Group & cluster with   & cluster without  \\
results  &        & cluster &   &  industry group   &  industry group \\
\hline
Mean return & &     &   &                   & \\
(2-stock) & 156.32 & 149.30 & 135.11 & 137.09 & 150.97 \\
(4-stock) & 152.99 & 151.27 & 134.02 & 132.85 & 155.49 \\
(8-stock) & 154.30 & 149.05 & 137.79 & 134.82 & 152.94 \\
\hline
Std. Dev. & &     &   &                   & \\
(2-stock) & 108.11 & 109.64 &  *99.57 & *66.02 & 116.40 \\
(4-stock) & 75.93  & 75.47  & *71.01  & *45.62 & 83.45 \\
(8-stock) & 51.34  & 50.15  & *48.26  & *32.49 & 57.63 \\
\hline
Sharpe Ratio & &     &   &                   & \\
(2-stock) & 1.45   & 1.36   & 1.36   & 2.07  & 1.27 \\
(4-stock) & 2.01   & 2.00   & 1.89   & 2.91  & 1.86 \\
(8-stock) & 3.01   & 2.97   & 2.86   & 4.15  & 2.65 \\
\hline
Levene Tests  & &     &   &                   & \\
(2-stock) & & & 0.05 & & $<10^{-16}$ \\
(4-stock) & & & 0.15 & & $<10^{-16}$ \\
(8-stock) & & & 0.06 & & $<10^{-16}$ \\
\hline
\end{tabular}
\caption{Period 3 portfolio performances with standard deviation for
three different portfolio sizes. The selection method with the lowest
standard deviation is marked with an asterisk (*).
The neighbor-Nets method is out-of-sample
testing using the correlation clusters determined from Period 2 data.
The first Levene test is the p-value whether the standard deviation of the
three different portfolio selection methods are equal. The second
result is for the neightbor-Net clusters split into dominant and
non-dominant industry groupings.} \label{tab:period3}
\end{table}

\begin{figure}[ht]
  \centering
  \includegraphics[width=14cm]{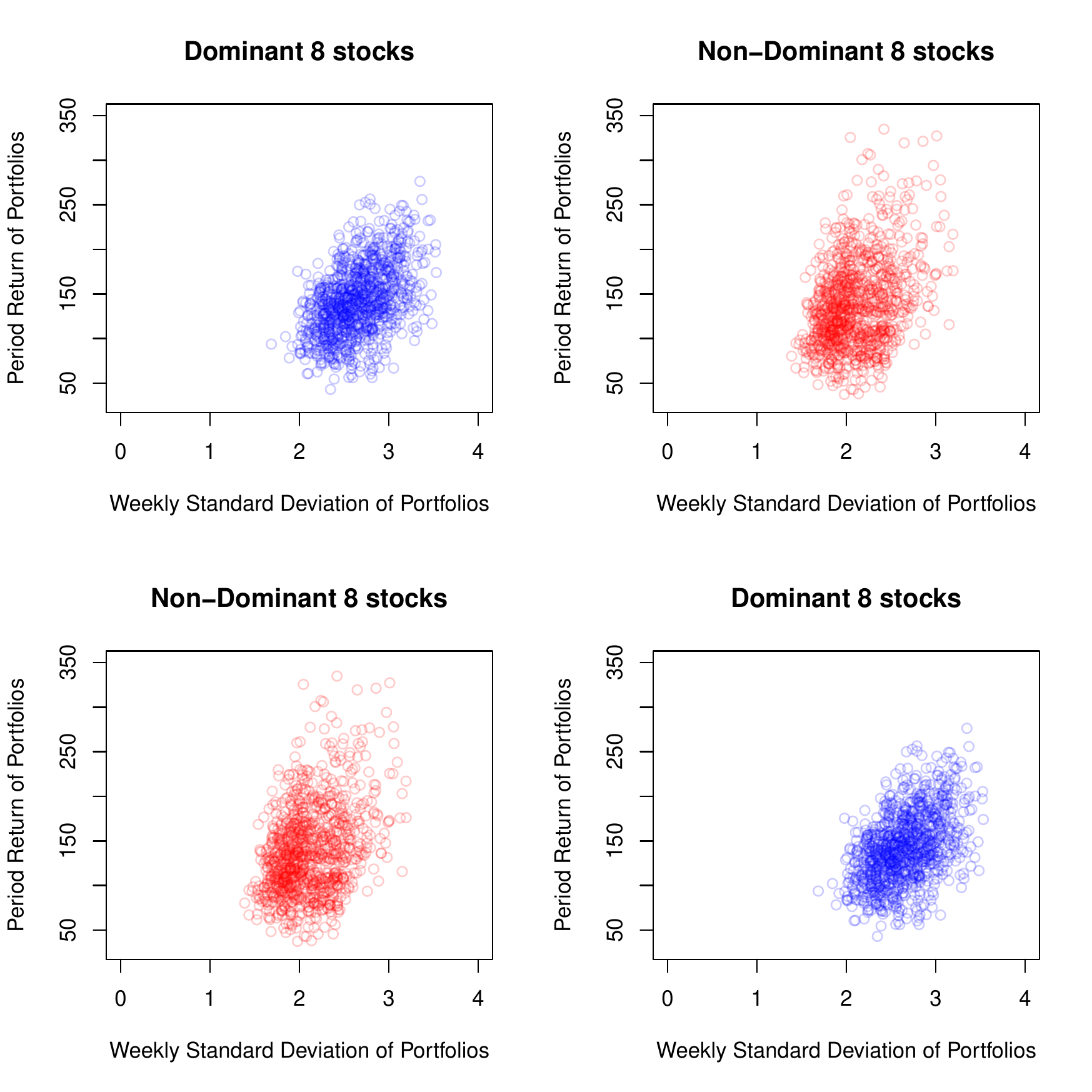}
  \caption{Scatter plots of Period 3 returns against the weekly
volatility for the eight-stock portfolios selected on the basis of
the clusters identified in  neighbor-Nets splits graph from
Period 2 with the clusters split into dominant and non-dominante
industry groups. }
\label{fig:P3Scatter8Stock}
\end{figure}

Table (\ref{tab:period3}) presents the results for the Period 3 simulations
which was a period of strongly rising equity prices. The model building
period (Period 2) was also a period of market increases.
Thus the out-of-sample test and model building periods
closely resemble each other.

The Levene tests show that only the two stock portfolios
had a significantly different spread of returns, though
the results for the eight-stock portfolios almost
reached statistical significance. The difference
between the neighbor-Nets selection and industry group selection
was almost 10 percentage points different.

The final two columns of Table (\ref{tab:period3}) present
the results for the simulations in which the correlation
clusters were divided into dominant and non-dominant 
industry groups. In this case the correlation clusters of
the dominant industries showed statistically significantly
lower levels of spread of the portfolio returns.

Figure (\ref{fig:P3Scatter8Stock}) plots the returns of
the weekly standard deviation of the portfolio returns,
a measure of volatility, against the period portfolio return for
the eight-stock portfolios dividing the stocks in to dominant
and non-dominant industry groups.
It is noticeable that the spread of returns
is smaller for the correlation clusters with dominant
industry groups though the weekly volatility is higher.

%
%



\begin{table}
\begin{tabular}{lrrr|rr}
Period 4 &         & Neighbor-Net’s & Industry & Correlation & Correlation  \\
Simulation & Random  & correlation 
 &  Group & cluster with   & cluster without  \\
results  &        & cluster &   &  industry group   &  industry group \\
\hline
Mean return & &     &   &                   & \\
(2-stock) & -46.58 & -49.18 & -45.56 & -54.61 & -43.26 \\
(4-stock) & -47.52 & -49.46 & -43.93 & -54.81 & -42.19 \\
(8-stock) & -47.48 & -48.52 & -44.09 & -54.70 & -42.18 \\
\hline
Std. Dev.  & &     &   &                   & \\
(2-stock) & 22.80 & *21.30 & 21.70 & *17.16 & 23.49 \\
(4-stock) & *15.16 & 15.41 & 15.68 & *11.29 & 16.80 \\
(8-stock) & 10.65 & 10.49 & *10.42 &  *8.03 & 11.11 \\
\hline
Sharpe Ratio  & &     &   &                   & \\
(2-stock) & -2.04 & -2.31 & -2.10  & -3.18 & -1.85 \\
(4-stock) & -3.13 & -3.21 & -2.80  & -4.85 & -2.51 \\
(8-stock) & -4.46 & -4.62 & -4.25  & -6.81 & -3.80 \\
\hline
Levene Tests  & &     &   &                   & \\
(2-stock) & &  & 0.78 & & $1.6\times 10^{-9}$ \\
(4-stock) & &  & 0.32 & & $<10^{-16}$          \\
(8-stock) & &  & 0.73 & & $<10^{-16}$          \\
\hline
\end{tabular}
\caption{Period 4 portfolio performances with standard deviation for
three different portfolio sizes. The selection method with the lowest
standard deviation is marked with an asterisk (*).
The neighbor-Nets method is out-of-sample
testing using the correlation clusters determined from Period 3 data.
The first Levene test is the p-value whether the standard deviation of the
three different portfolio selection methods are equal. The second
result is for the neightbor-Net clusters split into dominant and
non-dominant industry groupings.} \label{tab:period4}
\end{table}

\begin{figure}[ht]
  \centering
  \includegraphics[width=14cm]{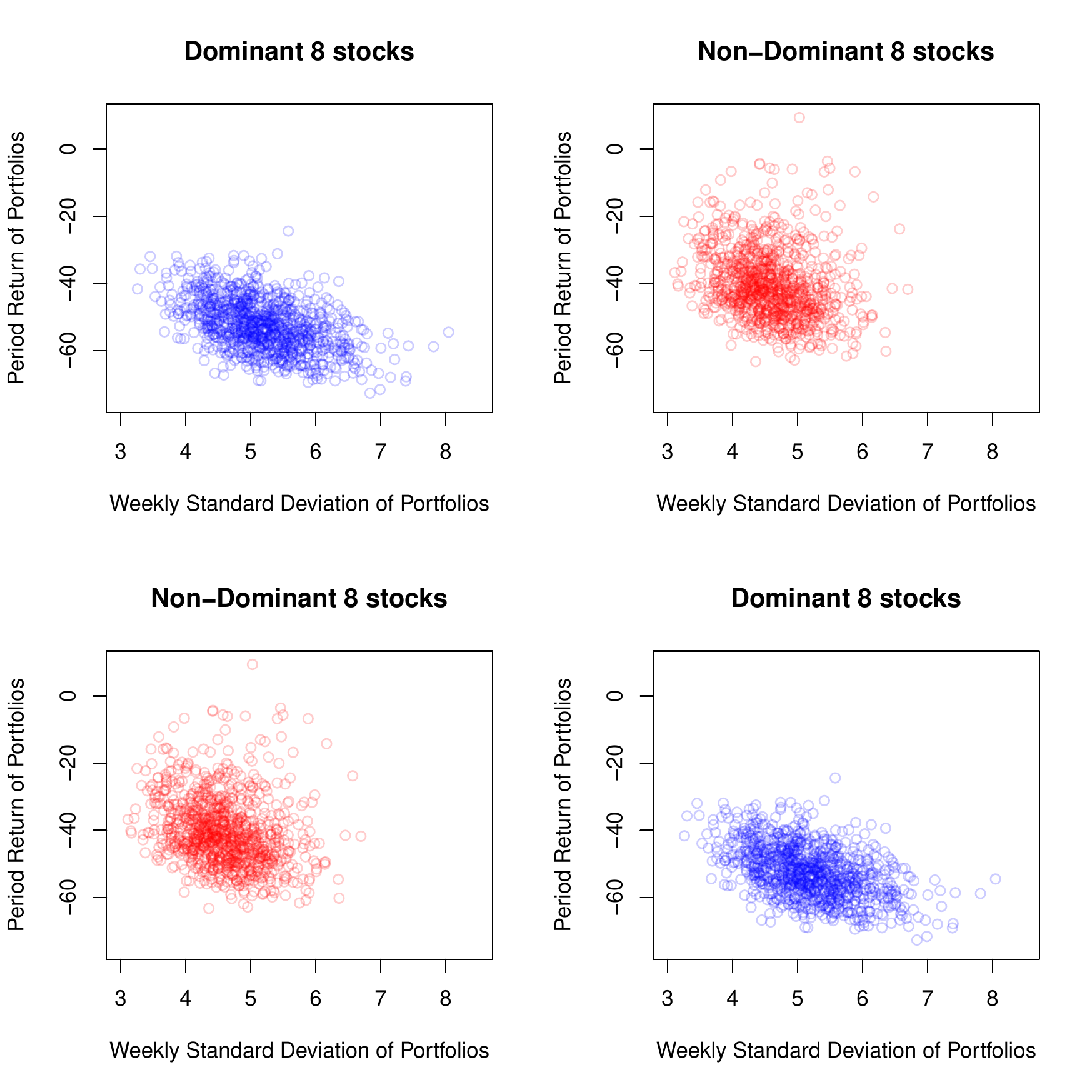}
  \caption{Scatter plots of Period 4 returns against the weekly
volatility for the eight-stock portfolios selected on the basis of
the clusters identified in  neighbor-Nets splits graph from
Period 3 with the clusters split into dominant and non-dominant
industry groups. }
\label{fig:P4Scatter8Stock}
\end{figure}

Table (\ref{tab:period4}) presents the result for the Period 4 simulations.
Period four was a period of strongly falling equity prices. The model building
period (Period 3) was a period of strong market increases.
Thus the out-of-sample test and model building periods
are effectively opposites of each other.

The Levene tests show that no stock portfolios
had a significantly different spread of returns. 

The final two columns of Table (\ref{tab:period4}) present
the results for the simulations in which the correlation
clusters were divided into dominant and non-dominant 
industry groups. In this case the correlation clusters of
the dominant industries showed statistically significantly
lower levels of spread of the portfolio returns. The Levene
tests were highly significant for all portfolio sizes.

Figure (\ref{fig:P4Scatter8Stock}) plots the returns of
the weekly standard deviation of the portfolio returns
against the period portfolio return for
the eight-stock portfolios dividing the stocks in to dominant
and non-dominant industry groups.
It is noticeable that the spread of returns
is substantially smaller for the correlation clusters with dominant
industry groups though, again, the weekly volatility is higher.

%
%


\begin{table}
\begin{tabular}{lrrr|rr}
Period 5 &         & Neighbor-Net’s & Industry & Correlation & Correlation  \\
Simulation & Random  & correlation 
 &  Group & cluster with   & cluster without  \\
results  &        & cluster &   &  industry group   &  industry group \\
\hline
Mean return & &     &   &                   & \\
(2-stock) & 164.49 & 159.20 & 156.22 & 142.77 & 160.07 \\
(4-stock) & 162.29 & 150.17 & 160.37 & 149.74 & 164.06 \\
(8-stock) & 162.43 & 154.90 & 162.25 & 151.25 & 163.34 \\
\hline
Std. Dev.  & &     &   &                   & \\
(2-stock)& 144.70 & 149.58 & *141.50 & *124.24 & 155.57 \\
(4-stock)& 105.14 &  *94.92 &  99.47 & *83.90  & 106.43\\
(8-stock)&  70.68 & *68.04  & 70.61  & *60.30  & 73.00 \\
\hline
Sharpe Ratio  & &     &   &                   & \\
(2-stock) & 1.14 & 1.06 & 1.10 & 1.15  & 1.03 \\
(4-stock) & 1.54 & 1.58 & 1.61 & 1.78  & 1.54 \\
(8-stock) & 2.30 & 2.28 & 2.38 & 2.51  & 2.24 \\
\hline
Levene Tests  & &     &   &                   & \\
(2-stock) & & & 0.46 & & $3.9\times 10^{-5}$ \\
(4-stock) & & & 0.003 & & $3.3\times 10^{-8}$ \\
(8-stock) & & & 0.35 & & $1.4\times 10^{-8}$ \\
\hline
\end{tabular}
\caption{Period 5 portfolio performances with standard deviation for
three different portfolio sizes. The selection method with the lowest
standard deviation is marked with an asterisk (*).
The neighbor-Nets method is out-of-sample
testing using the correlation clusters determined from Period 4 data.
The first Levene test is the p-value whether the standard deviation of the
three different portfolio selection methods are equal. The second
result is for the neightbor-Net clusters split into dominant and
non-dominant industry groupings.} \label{tab:period5}
\end{table}

\begin{figure}[ht]
  \centering
  \includegraphics[width=14cm]{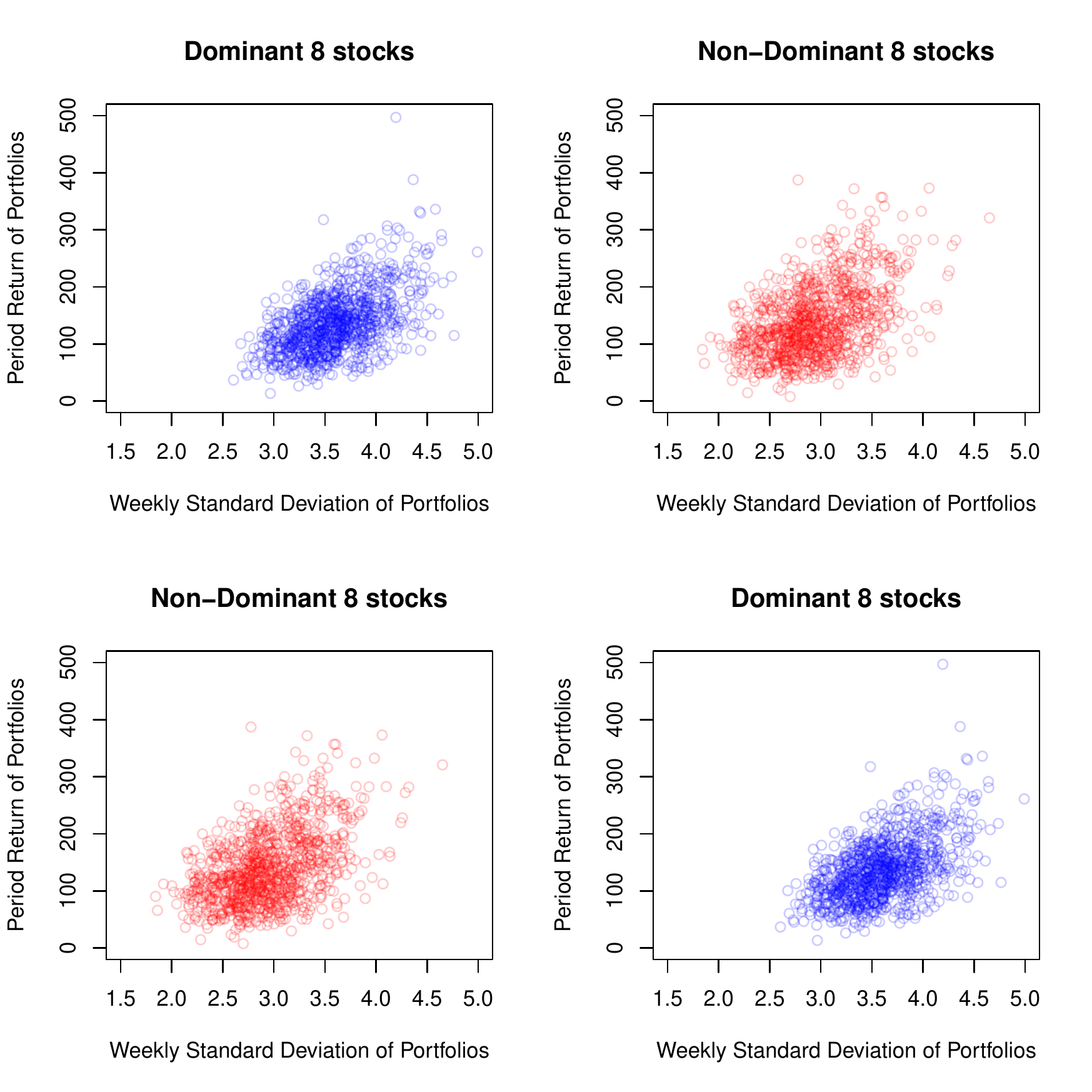}
  \caption{Scatter plots of Period 5 returns against the weekly
volatility for the eight-stock portfolios selected on the basis of
the clusters identified in  neighbor-Nets splits graph from
Period 4 with the clusters split into dominant and non-dominant
industry groups. }
\label{fig:P5Scatter8Stock}
\end{figure}

Table (\ref{tab:period5}) presents the results for the Period 5 simulations
which was a period of equity prices initially rebounding then
tracking sideways. The model building
period (Period 4) was a period of strong market decreases.
Thus the out-of-sample test and model building periods
are substantially different.

The Levene tests show that only the four stock portfolios
had a significantly different spread of returns. 

The final two columns of Table (\ref{tab:period5}) present
the results for the simulations in which the correlation
clusters were divided into dominant and non-dominant 
industry groups. In this case the correlation clusters of
the dominant industries showed statistically significantly
lower levels of spread of the portfolio returns. The Levene
tests were strongly significant for all portofilos sizes.

Figure (\ref{fig:P5Scatter8Stock}) plots the returns of
the weekly standard deviation of the portfolio returns
against the period portfolio return for
the eight-stock portfolios dividing the stocks in to dominant
and non-dominant industry groups.
It is noticeable that the spread of returns
is smaller for the correlation clusters with dominant
industry groups though, again, the weekly volatility is higher.

%
%


\begin{table}
\begin{tabular}{lrrr|rr}
Period 6 &         & Neighbor-Net’s & Industry & Correlation & Correlation  \\
Simulation & Random  & correlation
 &  Group & cluster with   & cluster without  \\
results  &        & cluster &   &  industry group   &  industry group \\
\hline
Mean return & &     &   &                   & \\
(2-stock)& 45.96 & 46.60 & 54.36 & 34.77 & 65.16 \\
(4-stock)& 48.39 & 51.10 & 55.19 & 35.85 & 61.87 \\
(8-stock)& 46.66 & 50.32 & 55.62 & 35.36 & 62.09 \\
\hline
Std. Dev.  & &     &   &                   & \\
(2-stock)& 52.03 & *49.87 & 51.62 & *44.92 & 51.32 \\
(4-stock)& 37.28 & *33.18 & 34.11 & *28.49 & 35.65 \\
(8-stock)& 25.56 & *21.63 & 22.94 & *15.04 & 24.04 \\
\hline
Sharpe Ratio  & &     &   &                   & \\
(2-stock)& 0.88 & 0.93 & 1.05 & 0.77  & 1.27 \\
(4-stock)& 1.30 & 1.54 & 1.62 & 1.26  & 1.74 \\
(8-stock)& 1.76 & 2.33 & 2.42 & 2.35  & 2.58 \\
\hline
Levene Tests  & &     &   &                   & \\
(2-stock) & &  & 0.64 & & 0.004 \\
(4-stock) & &  & 0.007 & & $4.0\times 10^{-9}$ \\
(8-stock) & &  & $1.1\times10^{-9}$ & & $<10^{-16}$          \\
\hline
\end{tabular}
\caption{Period 6 portfolio performances with standard deviation for
three different portfolio sizes. The selection method with the lowest
standard deviation is marked with an asterisk (*).
The neighbor-Nets method is out-of-sample
testing using the correlation clusters determined from Period 5 data.
The first Levene test is the p-value whether the standard deviation of the
three different portfolio selection methods are equal. The second
result is for the neightbor-Net clusters split into dominant and
non-dominant industry groupings.} \label{tab:period6}
\end{table}

Table (\ref{tab:period6}) presents the result for the Period 6 simulations
which was a period of rising equity prices with significant volatility. 
The model building
period (Period 5) was a period of rebound followed by a time
of stacking sideways.
Thus the out-of-sample test and model building periods
have some similarities.

The Levene tests show that the four  and eight stock portfolios
had a significantly different spread of returns.  In both cases
the neighbor-Nets portfolio selection method had the lowest
spread of returns.

The final two columns of Table (\ref{tab:period6}) present
the results for the simulations in which the correlation
clusters were divided into dominant and non-dominant 
industry groups. In this case the correlation clusters of
the dominant industries showed statistically significantly
lower levels of spread of the portfolio returns. The Levene
tests were highly significant for all portofilos sizes.

Figure (\ref{fig:P6Scatter8Stock}) plots the returns of
the weekly standard deviation of the portfolio returns
against the period portfolio return for
the eight-stock portfolios dividing the stocks in to dominant
and non-dominant industry groups.
It is noticeable that the spread of returns
is smaller for the correlation clusters with dominant
industry groups though, as with some previous periods, 
the weekly volatility is higher.

\section{Discussion and Conclusions}\label{sec:Discuss}

The simulation tests performed here represent a particularly severe
test of portfolio diversification because of the long out-of-sample
test periods coupled with the fact that the market conditions in the 
model building phase were often very different from the market
conditions in the test phase. 

In the 15 sets of simulations comparing, random, industry group,
and neighbor-Net correlation cluster selection methods,
 statistically significant differences
in the portfolios standard deviation were obtained
only five times. In four cases 
the neighbor-Net correlation cluster produced the lowest standard deviation
and in one case the industry group selection method was the lowest.

Considering the neighbor-Net correlation clusters split into dominant
and non-dominant industries, all 15 cases had statistically significant 
differences in the portfolios' standard deviation. Of these 12 were
correlation clusters with dominant industry groups in Periods 3
through 6 and three
were correlation clusters with non-dominant industry groups in Period 2.
Because we do not have data from before the inception date of the ASX200
index it is not possible to say if this represents a change in the behaviour
of the market. 

Figures (\ref{fig:P2Scatter8Stock}) through (\ref{fig:P6Scatter8Stock})
present this last observation in graphical form and it is clear that
the distribution both in terms of portfolio returns and weekly volatility 
are different for all periods. Graphs of the two and four stock
portfolios show similar results but are not reported here.

These results suggest that within each correlation cluster there are
two distinct sub-populations of stocks. Intuitively, the dominant 
industry group are simply stocks within the same industry with similar
risk characteristics. We would expect such stocks to be strongly correlated,
hence fall into the same correlation cluster. We would also expect that
they would continue to be strongly correlated into the future. 
On the other hand, the stocks in the non-dominant industry group within
a correlation cluster would seem far less likely to remain strongly
correlated in the future. 

At this stage we can only recommend that this be investigated further. 
The differences between the two groups of stocks appear significant both
in a statistical and economic sense but it is not yet clear how to
exploit this difference for financial gain.

This paper is primarily concerned about reducing risk in small
portfolios. However, investors are also concerned about returns
and in three of the five periods (two, three and five) 
the randomly selected portfolios
had the highest returns among the three methods which included
all stocks. Truly negative correlations among stocks are uncommon
and in a strongly rising market a negative correlation between
a pair of stocks often indicates that one stock is suffering from
some form of financial distress and hence a falling share price.
The correlation cluster method excludes many stock combinations
available to the random selection method but also increases the
probability of a stock being paired with one in financial distress
hence depressing overall portfolio performance. 
If this explanation is correct then the application of financial
analysis aimed at removing financially distressed stocks may well
enhance the performance of all portfolio selection methods. Again,
this must wait for further research.

\bibliography{StockNNet}

\appendix

\section{Neighbor-Net}\label{app:neighbornet}

This appendix gives a more technical description of the Neighbor-Net algorithm.

The construction of Neighbor-Net networks  has four key components: the agglomerative process, selection formulae, distance reduction and estimation of the split weights. The agglomerative process describes how the hierarchy of nodes is determined, selection formulae describe the system used in determining the hierarchy and distance reduction describes how the distances are adjusted as the hierarchy is built. The result of the these three steps is a circular collection of splits. Formally a set of circular splits is one which satisfies that condition that there is an ordering of the nodes $x_1,x_2, \ldots, x_n$ such that every split is of the form $\{x_i, x_{i+1},\ldots, x_j \}| X- \{x_i,\ldots, x_j \}$ for some $i$ and $j$ satisfying $1 \le i \le j <n$. As highlighted above, the advantage of this set of splits is that they can be represented on a plane.
 
We describe the algorithm following \cite{Bryant2004}. All the nodes start out as singletons and the selection formulae finds the two closest nodes. These nodes are not grouped immediately but remain as singletons until a node has two neighbors. At this stage the three nodes, the node and its two neighbors, are merged into two nodes. Here we present the selection formula for grouping nodes. Let neighboring relations group the $n$ nodes into $m$ clusters. Let $d_{xy}$ be the distance between nodes $x$ and $y$.  Let $C_1, C_2, \ldots, C_m, m \le n$ be the $m$ clusters. The distance $d(C_i, C_j)$ between two clusters is
\begin{equation}
d(C_i, C_j) = \frac{1}{|C_i| |C_j|} \sum_{x \in C_i} \sum_{y \in C_j} d_{xy},
\end{equation}
that is, an average of the distances between elements in each cluster. 

The closest pair of clusters is given by finding the $i$ and $j$ that minimise 
\begin{equation}
Q (C_i, C_j) = (m-2)d(C_i, C_j) - \sum_{k=i, k \ne i}{m} d(C_i, C_k) -  \sum_{k=i, k \ne i}{m} d(C_j, C_k),
\end{equation}
and denote them $C_{i*}$ and $C_{j*}$

To choose particular nodes within clusters we select the node from each cluster that minimises
\begin{equation}
\hat{Q} (x_i, x_j) = (\hat{m}-2)d(x_i, x_j) - \sum_{k=i, k \ne i}{\hat{m}} d(x_i, C_k) -  \sum_{k=i, k \ne i}{\hat{m}} d(x_j, C_k) 
\end{equation}
where $x_i \in C_{i*}$ and $x_j \in C_{j*}$ and $\hat{m} = m + |C_{i*}| + |C_{j*}| - 2$.

The distance reduction updates the distance matrix with the distance from the two new clusters to all the other clusters. The distance reduction formulae calculate the distances between the existing nodes and the new combined nodes. If $y$ has two neighbors, $x$ and $z$, then the three nodes will be combined and replaced by two nodes which we can denote as $u$ and $v$. The Neighbor-Net algorithm uses
\begin{eqnarray}
d(u, a) &=& (\alpha + \beta)d(x,a) + \gamma d(y,a) \\
d(v, a) &=& \alpha d(y,a) + (\beta + \gamma) d(z,a) \\
d(u, v) &=& \alpha d(x,y) + \beta d(x,z) + \gamma d(y,z)
\end{eqnarray}
where $\alpha, \beta$ and $\gamma$ are non-negative real numbers with $\alpha + \beta + \gamma = 1$.

The process stops when all the nodes are in a single cluster. 

The Neighbor-Net method of \cite{Bryant2004} used non-negative least squares to estimate the split weights given the distance vector and a set splits known as the circular splits. Suppose that the splits in the network are numbered $1,2, \ldots, m$ and that the nodes are numbered $1,2, \ldots, n$. Let $\mathbf{X}$ be the be the {\em splits matrix} with the dimensions $n(n-1)/2 \times m$ matrix with rows indexed by pairs of nodes, columns indexed by splits, and entry $\mathbf{X}_{ij,k}$ given by
\begin{eqnarray}
\mathbf{X}_{ij,k} &=& 
\begin{cases}
1 \text{ if $i$ and $j$ are on opposite sides of the split} \\
0 \text{ if $i$ and $j$ are on the same side of the split.}
\end{cases}
\end{eqnarray}

Similar nodes will be clustered together in the network. This is a direct result of each pair of neighboring nodes in the ordering being close together in terms of distance, and separated from node where the distance measure reveals dissimilarity. 

The network, or splits graph, generated by Neighbor-Nets has
three biologically meaningful components. The places where a line
splits represents a speciation event, where a single population
becomes two genetically isolated populations. The places where
two lines join to become one represents a recombination event,
where two genetically isolated populations exchange genetic
material. The lengths of the indivdual lines represent the
time the population evolves without either a speciation
or recombination event. The interpretation of these three
components in a financial context is an active area of 
research for the authors.

\end{document}